\documentclass[submission, PhysProc]{SciPost}

\hypersetup{
    colorlinks,
    linkcolor={red!50!black},
    citecolor={blue!50!black},
    urlcolor={blue!80!black}
}

\usepackage[bitstream-charter]{mathdesign}
\DeclareSymbolFont{usualmathcal}{OMS}{cmsy}{m}{n}
\DeclareSymbolFontAlphabet{\mathcal}{usualmathcal}

\begin{document}

\begin{center}{\Large \textbf{
Four-Body Faddeev-Type Calculation of the $\bar{K}NNN$ System: Preliminary Results\\
}}\end{center}

% TODO: write the author list here. Use first name (+ other initials) + surname format.
% Separate subsequent authors by a comma, omit comma and use "and" for the last author.
% Mark the corresponding author with a superscript star.
\begin{center}
N. V. Shevchenko\textsuperscript{1$\star$}
\end{center}

\begin{center}
{\bf 1} Nuclear Physics Institute of the CAS, \v{R}e\v{z} 250 68, Czech Republic
\\
${}^\star$ {\small \sf shevchenko@ujf.cas.cz}
\end{center}

\begin{center}
\today
\end{center}

% For convenience during refereeing (optional),
% you can turn on line numbers by uncommenting the next line:
%\linenumbers

\definecolor{palegray}{gray}{0.95}
\begin{center}
\colorbox{palegray}{
  \begin{tabular}{rr}
  \begin{minipage}{0.05\textwidth}
    \includegraphics[width=14mm]{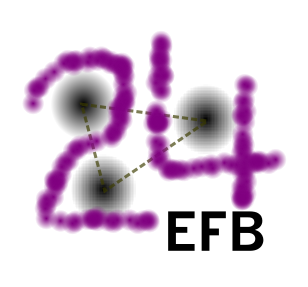}
  \end{minipage}
  &
  \begin{minipage}{0.82\textwidth}
    \begin{center}
    {\it Proceedings for the 24th edition of European Few Body Conference,}\\
    {\it Surrey, UK, 2-4 September 2019} \\
    %\doi{10.21468/SciPostPhysProc.2}\\
    \end{center}
  \end{minipage}
\end{tabular}
}
\end{center}

\section*{Abstract}
{\bf
% TODO: write your abstract here.
The paper is devoted to the $\bar{K}NNN$ system, consisting of an antikaon and three nucleons.
Four-body Faddeev-type AGS equations are being solved in order to find possible quasi-bound state
in the system.
}

\vspace{10pt}
\noindent\rule{\textwidth}{1pt}
\tableofcontents\thispagestyle{fancy}
\noindent\rule{\textwidth}{1pt}
\vspace{10pt}

%%%%%%%%%%%%%%%%%%%%%%%%%%%%%%%%%%%%%%%%%%%%%%%%%%%%%%%%%%%%

\section{Introduction}

Attractive nature of $\bar{K}N$ integration lead to suggestions, that quasi-bound states
can exist in few-body systems consisting of antikaons and nucleons~\cite{AY1}.
In particular, a deep and relatively narrow quasi-bound state was predicted  in the
lightest three-body $\bar{K}NN$ system~\cite{AY2}. Many theoretical calculations of the system
were performed after that using different methods and inputs, see e.g. \cite{review}. 
All of them agree, that the quasi-bound state really exists in spin-zero state of $\bar{K}NN$,
usually denoted as $K^- pp$, but predict quite different binding energies and widths of the state.

In recent years we performed a series of calculations of different properties and states of the
three-body $\bar{K}NN$ and $\bar{K}\bar{K}N$ systems \cite{review}, using dynamically exact
Faddeev-type equations in AGS form with coupled $\bar{K}NN$ and $\pi \Sigma N$ channels.
In particular, we predicted $K^- pp$ quasi-bound state binding energy and width using three
different models of $\bar{K}N$ interaction. The same was done for the $\bar{K}\bar{K}N$ system.
We also demonstrated, that there is no quasi-bound states, caused by pure strong interactions,
in another spin state of $\bar{K}NN$ system, which is $K^- d$. In addition, we calculated
the near-threshold amplitudes of $K^-$ elastic scattering on deuteron.  Finally, we evaluated $1s$
level shift in kaonic deuterium, which is an atomic state, caused by presence of the strong
$\bar{K}N$ interaction in comparison to the pure Coulomb state.

The next step is study of a four-body $\bar{K}NNN$ system. Some calculations were already
performed for it~\cite{AY1,BGL,ir}, but more accurate calculations are needed. We use four-body
Faddeev-type equations in AGS form~\cite{4AGS}. Only these dynamically exact equations in
momentum re\-pre\-sen\-tation can treat energy-dependent $\bar{K}N$ potentials, necessary
for the this system, exactly.

\section{Four-body Faddeev-type AGS equations for the $\bar{K}NNN$ system}

Description of the four-body Faddeev-type equations in AGS form~\cite{4AGS} should start
from the three-body AGS equations~\cite{AGS} since the three-body transition operators,
being solution of the three-body equations, enter the four-body ones together with two-body
$T$-matrices. If separable potentials of the form
$V_{\alpha} = \lambda_{\alpha} |g_{\alpha} \rangle \langle g_{\alpha}|$ are used,
the three-body transitions operators  satisfy the three-body Faddeev-type
equations in AGS form~\cite{AGS}:
\begin{equation}
\label{3AGSsep}
 X_{\alpha \beta}(z) = Z_{\alpha \beta}(z) + 
 \sum_{\gamma=1}^3 Z_{\alpha \gamma}(z) \tau_{\gamma}(z) X_{\gamma \beta}(z)
\end{equation}
with transition $X_{\alpha \beta}$ and kernel $Z_{\alpha \beta}$ operators are defined as
\begin{eqnarray}
\label{X3b}
 X_{\alpha \beta}(z) &=&
  \langle g_{\alpha} | G_0(z) U_{\alpha \beta}(z) G_0(z) | g_{\beta} \rangle, \\
\label{Z3b}
 Z_{\alpha \beta}(z) &=& (1-\delta_{\alpha \beta})
   \langle g_{\alpha} | G_0(z) | g_{\beta} \rangle.
\end{eqnarray}
Operator $U_{\alpha \beta}(z)$ in Eq.(\ref{X3b}) is a three-body transition operator
of the general form, which describes process
$\beta + (\alpha \gamma) \to \alpha + (\beta \gamma)$, while $G_0(z)$ in Eqs.(\ref{X3b},\ref{Z3b})
is the three-body free Green function. Faddeev partition indices $\alpha, \beta = 1,2,3$
simultaneously define a particle ($\alpha$) and the remained pair ($\beta \gamma$).
The operator $\tau_{\alpha}(z)$ in Eq.(\ref{3AGSsep}) is an energy-dependent part of
a separable two-body $T$-matrix
$
T_{\alpha}(z) = |g_{\alpha} \rangle \tau_{\alpha}(z) \langle g_{\alpha}|,
$
corresponding to the separable potential describing interaction in the ($\beta \gamma$)
pair; $| g_{\alpha} \rangle$ is a form-factor.

%------------------------------------------------------------------------------
The four-body Faddeev-type AGS equations~\cite{4AGS}, written for separable potentials, have a form
\begin{eqnarray}
\label{4AGSsepV}
 \bar{U}^{\sigma \rho}_{\alpha \beta}(z) &=& (1-\delta_{\sigma \rho}) 
  (\bar{G_0}^{-1})_{\alpha \beta}(z) + 
 \sum_{\tau,\gamma,\delta} (1-\delta_{\sigma \tau}) \bar{T}^{\tau}_{\alpha \gamma}(z)
  (\bar{G_0})_{\gamma \delta}(z) \bar{U}^{\tau \rho}_{\delta \beta}(z), \\
\label{barU}
 &{}& \bar{U}^{\sigma \rho}_{\alpha \beta}(z) = 
  \langle g_{\alpha} | G_0(z) U^{\sigma \rho}_{\alpha \beta}(z) G_0(z) | g_{\beta} \rangle, \\
\label{barT}
  &{}& \bar{T}^{\tau}_{\alpha \beta}(z) = 
  \langle g_{\alpha} | G_0(z) U^{\tau}_{\alpha \beta}(z) G_0(z) | g_{\beta} \rangle, 
\; \\
\label{barG0}
 &{}& (\bar{G_0})_{\alpha \beta}(z) = \delta_{\alpha \beta} \tau_{\alpha}(z).
\end{eqnarray}
Here operators $\bar{U}^{\sigma \rho}_{\alpha \beta}$ and $\bar{T}^{\tau}_{\alpha \beta}$
contain four-body $U_{\alpha \beta}^{\sigma \rho}(z)$ and three-body $U_{\alpha \beta}^{\tau}(z)$
transition operators of the general form, correspondingly. The lower indices $\alpha,\beta$ in
Eqs.(\ref{4AGSsepV},\ref{barU},\ref{barT}) similarly to the case of the three-body equations
(\ref{3AGSsep},\ref{X3b},\ref{Z3b}) define two-body subsystems of the full system. The upper
indices $\tau,\sigma,\rho$ define a partition of the four-body system which can be of $3+1$ or
$2+2$ type. In particular, there are two partitions of $3+1$ type: $|\bar{K} + (NNN) \rangle$,
$|N + (\bar{K}NN) \rangle$, - and one of the $2+2$ type: $|(\bar{K}N) + (NN) \rangle$, -
for the $\bar{K}NNN$ system. The free Green function $G_0(z)$ acts in four-body space.

The four-body system of AGS equations Eq.(\ref{4AGSsepV}) look similar to the three-body AGS
system with arbitrary potentials. If we, as suggested in \cite{4AGSsep}, represent
the ''effective three-body poten\-tials'' $\bar{T}^{\tau}_{\alpha \beta}(z)$ in Eq.~(\ref{4AGSsepV})
in a separable form
$
\bar{T}^{\tau}_{\alpha \beta}(z) =  | \bar{g}^{\tau}_{\alpha} \rangle 
  \bar{\tau}^{\tau}_{\alpha \beta}(z)   \langle \bar{g}^{\tau}_{\alpha} |,
$
the four-body equations can be rewritten as \cite{4AGSsep}
\begin{equation}
\label{4AGSsepVT}
 \bar{X}^{\sigma \rho}_{\alpha \beta}(z) = \bar{Z}^{\sigma \rho}_{\alpha \beta}(z) + 
 \sum_{\tau,\gamma,\delta} \bar{Z}^{\sigma \tau}_{\alpha \gamma}(z) \bar{\tau}^{\tau}_{\gamma \delta}(z)
   \bar{X}^{\tau \rho}_{\delta \beta}(z)
\end{equation}
with new four-body transition $\bar{X}^{\sigma \rho}$ and kernel $\bar{Z}^{\sigma \rho}$ operators
\begin{eqnarray}
 \bar{X}^{\sigma \rho}_{\alpha \beta}(z) &=&
  \langle \bar{g}^{\sigma}_{\alpha} | \bar{G_0}(z)_{\alpha \alpha} \bar{U}^{\sigma \rho}_{\alpha \beta}(z)
    \bar{G_0}(z)_{\beta \beta} | \bar{g}^{\rho}_{\beta} \rangle, \\
 \bar{Z}^{\sigma \rho}_{\alpha \beta}(z) &=& (1-\delta_{\sigma \rho})
   \langle \bar{g}^{\sigma}_{\alpha} | \bar{G_0}(z)_{\alpha \beta} | \bar{g}^{\rho}_{\beta} \rangle.
\end{eqnarray}

%%%%%%%%%%%%%%%%%%%%%%%%%%%%%%%%%%%%%%%%%%%%%%%%%%%%%%%%%%
Necessary for the $\bar{K}NNN$ calculations $\bar{K}N$ and $NN$ potentials, which we use, are separable
ones by construction.
%Therefore, we need to separabelize only three-body and 2+2 amplitudes, entering the equations~(\ref{4AGSsepVT}).
Therefore, we need to construct separable versions of three-body and 2+2 amplitudes only,
entering the equations~(\ref{4AGSsepVT}).
We  use Energy Dependent Pole Expansion/ Approximation
(EDPE/ EDPA) method, suggested in~\cite{EDPE} specially for the four-body AGS equa\-tions.

Three-body Faddeev-type AGS equations written in momentum basis 
for $s$-wave interactions have a form:
\begin{equation}
\label{3AGS}
 X_{\alpha \beta}(p,p';z) = Z_{\alpha \beta}(p,p';z) +
 \sum_{\gamma = 1}^3 4 \pi
 \int_{0}^{\infty} Z_{\alpha \gamma}(p,p'';z) \, \tau_{\gamma}(p'';z) \, X_{\gamma \beta}(p'',p';z) p''^2 dp'',
\end{equation}
were $p,p'$ and $z$ are three-body momenta and energy.
Eigenvalues $\lambda_n$ and eigenfunctions $g_{n \alpha}(p;z)$ of the system Eq.(\ref{3AGS}) can be evaluated
from the system of equations
\begin{equation}
\label{gHS}
 g_{n \alpha}(p;z) = \frac{1}{\lambda_n} \,
 \sum_{\gamma = 1}^3 4 \pi
 \int_{0}^{\infty} Z_{\alpha \gamma}(p,p';z) \, \tau_{\gamma}(p';z) \, g_{n \gamma}(p';z) p'^2 dp'
\end{equation}
with normalization condition
\begin{equation}
\label{normHS}
 \sum_{\gamma = 1}^3 4 \pi
 \int_{0}^{\infty} g_{n \gamma}(p';z) \, \tau_{\gamma}(p';z) \, g_{n' \gamma}(p';z) p'^2 dp' = -\delta_{nn'}.
\end{equation}

EDPE/EDPA method needs solution of the eigenequations~Eq.(\ref{gHS}) for a fixed energy $z$, which
usually is chosen to be the binding energy $z=E_B$. After that energy dependent form-factors
\begin{equation}
 \label{gEDPE}
 g_{n \alpha}(p;z) = \frac{1}{\lambda_n} \,
 \sum_{\gamma = 1}^3 4 \pi
 \int_{0}^{\infty} Z_{\alpha \gamma}(p,p';z) \, \tau_{\gamma}(p';E_B) \, g_{n \gamma}(p';E_B) p'^2 dp'
\end{equation}
and propagators
\begin{eqnarray}
\label{thetaEDPE}
 \left( \Theta(z) \right)^{-1}_{mn} \, = \,
 \sum_{\gamma = 1}^3 4 \pi
 \int_{0}^{\infty} g_{m \gamma}(p';z) \, \tau_{\gamma}(p';E_B) \, g_{n \gamma}(p';E_B) p'^2 dp'  \\
\nonumber
-  \sum_{\gamma = 1}^3 4 \pi
 \int_{0}^{\infty} g_{m \gamma}(p';z) \, \tau_{\gamma}(p';z) \, g_{n \gamma}(p';z) p'^2 dp'
\end{eqnarray}
can be calculated. 
%Finally, the separabelized three-body amplitude has a form
Finally, the separable version of a  three-body amplitude has a form
\begin{equation}
\label{X_EDPE}
 X_{\alpha \beta}(p,p';z) = \sum_{m,n=1}^{\infty}  g_{m \alpha}(p;z) \, \Theta_{mn}(z) \, g_{n \beta}(p';z).
\end{equation}
If only one term is taken in the sums in~Eq.(\ref{X_EDPE}), the Energy Dependent Pole Expansion
turns into Energy Dependent Pole Approximation. It is seen, that in contrast to Hilbert-Schmidt expansion,
EDPE method needs only one solution of the eigenvalue equations Eq.(\ref{gHS}) and calculations of the
integrals Eqs.(\ref{gEDPE},\ref{thetaEDPE}) after that. According to the authors, the method is accurate
already with one term (i.e. EDPA), and it converges faster than Hilbert-Schmidt expansion.

%%%%%%%%%%%%%%%%%%%%%%%%%%%%%%%%%%%%%%%%%%%%%%%%%%%%%%%%%%
In order to find the quasi-bound state energy the homogeneous system of equations Eq.(\ref{4AGSsepVT})
has to be solved. We started by writing down the system Eq.(\ref{4AGSsepVT}) for $18$ channels
$\sigma_{\alpha}$ with $\alpha =$ $NN$ or $\bar{K}N$, considering three nucleons as nonindentical particles:
%At the begin we considered all three nucleons as different particles, so we started by writing
%down the homogeneous system of equations Eq.(\ref{4AGSsepVT}) for the following $18$ channels
%$\sigma_{\alpha}$ with $\alpha =$ $NN$ or $\bar{K}N$:
\begin{eqnarray}
\nonumber
&{}&  1_{NN}  :  |\bar{K} + (N_1 + N_2 N_3) \rangle, |\bar{K} + (N_2 + N_3 N_1) \rangle,
                      |\bar{K} + (N_3 + N_1 N_2) \rangle,          \\
\nonumber
&{}&  2_{NN}:  |N_1 + (\bar{K} + N_2 N_3) \rangle, |N_2 + (\bar{K} + N_3 N_1) \rangle,
                      |N_3 + (\bar{K} + N_1 N_2) \rangle, \\
\label{channels}                      
&{}&  2_{\bar{K}N}: |N_1 + (N_2 + \bar{K} N_3) \rangle, |N_2 + (N_3 + \bar{K} N_1) \rangle,
                      |N_3 + (N_1 + \bar{K} N_2) \rangle,\\
\nonumber                      
&{}&  \quad \quad \quad \! |N_1 + (N_3 + \bar{K} N_2) \rangle, |N_2 + (N_1 + \bar{K} N_3) \rangle,
                      |N_3 + (N_2 + \bar{K} N_1) \rangle,\\
\nonumber
&{}&  3_{NN}:  |(N_2 N_3) + (\bar{K} + N_1) \rangle, |(N_3 N_1) + (\bar{K} + N_2) \rangle,
                   |(N_1 N_2) + (\bar{K} + N_3) \rangle, \\
\nonumber                  
&{}&  3_{\bar{K}N}: |(\bar{K} N_1) + (N_2 + N_3) \rangle, |(\bar{K} N_2) + (N_3 + N_1) \rangle,
                      |(\bar{K} N_3) + (N_1 + N_2) \rangle
\end{eqnarray}
After antisymmetrization, necessary for a system with identical fermions, the system of equations
to be solved can be written in a matrix form:
\begin{equation}
\label{4AGSfinal}
 \hat{X} = \hat{Z} \, \hat{\tau} \, \hat{X}.
\end{equation}
 
Our $\bar{K}N$ and $NN$ potentials, which we use for the $\bar{K}NNN$ system calculations, are isospin-
and spin-dependent ones. In addition, our $NN$ interaction model is a two-term potential. 
%At the first step
%only one term was taken in separabelization of the three-body $\bar{K}NN$, $NNN$ and 2+2 $\bar{K}N + NN$
%amplitudes in Eq. (\ref{X_EDPE}) (EDPA).
At the first step only one separable term was used for the three-body $\bar{K}NN$, $NNN$ and
2+2 $\bar{K}N + NN$ amplitudes in Eq. (\ref{X_EDPE}) (EDPA).
Keeping all this in mind, matrices $\hat{Z}$ and $\hat{\tau}$,
entering the antisymmetrized equations Eq.(\ref{4AGSfinal}) are matrices $18x18$ containing the kernel
operators $\bar{Z}^{\sigma \rho}_{\alpha}$ and $\bar{\tau}^{\rho}_{\alpha \beta}$, correspondingly.

\section{Two-body potentials, $3+1$ and $2+2$ partitions}

\subsection{Two-body potentials}

Both $\bar{K}N$ and $NN$ potentials are separable isospin- and spin-dependent ones in $s$-wave.
We use three our separable antikaon-nucleon potentials constructed for our three-body calculations
of the $\bar{K}NN$ and $\bar{K} \bar{K} N$ systems. They are:  two phenomenological potentials with
coupled $\bar{K}N - \pi \Sigma$ channels, having one- or two-pole structure of the $\Lambda(1405)$
resonance~\cite{NPA890} and a chirally motivated model with coupled $\bar{K}N - \pi \Sigma - \pi \Lambda$
channels and two-pole structure~\cite{PRC90-I}. All three potentials describe low-energy $K^- p$ scattering,
namely: elastic
$K^- p \to K^- p$ and inelastic $K^- p \to MB$ cross-sections and threshold branching ratios $\gamma, R_c, R_n$.
They also reproduce $1s$ level shift of kaonic hydrogen caused by the strong $\bar{K}N$ interaction
in comparison to the pure Coulomb level, measured by SIDDHARTA experiment~\cite{SIDD}: 
$\Delta_{1s}^{SIDD} = -283 \pm 36 \pm 6$ eV, $\Gamma_{1s}^{SIDD} = 541 \pm 89 \pm 22$ eV. All
the experimental data are described by three our potentials with equally high accuracy. In addition,
elastic $\pi \Sigma$ cross-sections with isospin $I_{\pi \Sigma}$ provided by all three potentials have a bump
in a region of the $\Lambda(1405)$ resonance (according to PDG~\cite{PDG}:
$M_{\Lambda(1405)}^{PDG} = 1405.1^{+1.3}_{-1.0}$ MeV, $\Gamma_{\Lambda(1405)}^{PDG} = 50.5 \pm 2.0$ MeV).
The poles corresponding to the $\Lambda(1405)$ resonance are situated at
\begin{eqnarray}
\label{zLambda1405}
z_{\Lambda(1405)-1}^{\rm SIDD1} &=& 1426 - i \, 48 {\, \rm MeV} \\
z_{\Lambda(1405)-1}^{\rm SIDD2} &=& 1414 - i \, 58 {\, \rm MeV}, \quad
 z_{\Lambda(1405)-2}^{\rm SIDD2} = 1386 - i \, 104 {\, \rm MeV}
\end{eqnarray}
for the phenomenological potentials with one- and two-pole structure, correspondingly~\cite{PRC90-I}, and at
\begin{equation}
z_{\Lambda(1405)-1}^{\rm Chiral} = 1417 - i \, 33 {\, \rm MeV}, \quad
 z_{\Lambda(1405)-2}^{\rm Chiral} = 1406 - i \, 89 {\, \rm MeV}
\end{equation}
for the chirally motivated potential~\cite{PRC90-II}.

The three antikaon-nucleon potentials with coupled $\bar{K}N - \pi \Sigma$ channels were used in three-body
AGS equations with coupled $\bar{K}NN - \pi \Sigma N$ three-body channels. By this the channel coupling was
taken into account in a direct way. The four-body AGS equations, which we solve, are too complicated to do
the same. Due to this we use the exact optical versions of our $\bar{K}N$ potentials. They have exactly the same
elastic part of the potential as the potential with coupled channels, while all in-elasticity is taken into account in
an energy-dependent imaginary part of the potential. It was demonstrated in our three-body calculations,
that such potentials give very accurate results in comparison with the results obtained with the coupled-channel
potentials, see below. Due to this we assume that it is a good approximation for the four-body calculations
as well.

We constructed a new version of the two-term separable $NN$ potential. It reproduces Argonne v18 $NN$ phase
shifts at low energies up to $500$ MeV with change of sign, which means it is repulsive at short distances.
It provides the following singlet and triplet $NN$ scattering lengths: $a_{s} = 16.32$ fm, $a_{t} = -5.40$ fm, and give
the deuteron binding energy $E_{deu} = 2.225$ MeV.

No three-body potentials were used since the four-body Faddeev-type equations are too complicated
in their original form with "normal" pair potentials already.

\subsection{$3+1$ and $2+2$ partitions}

We are studying the $\bar{K}NNN$ system with the lowest value of the four-body isospin $I^{(4)}=0$, which can be
denoted as $K^- ppn$. Its total spin $S^{(4)}$ is equal to one half, while the orbital momentum is zero, since all
two-body interactions are chosen to be $s$-wave ones. For the $\bar{K}NNN$ system with these quantum numbers
the following three-body subsystems contribute:
\begin{itemize}
 \item $\bar{K}NN$ with  isospin $I^{(3)} = 1/2$ and spin $S^{(3)} = 0$ ($K^- pp$) or spin $S^{(3)} = 1$ ($K^- d$).
 \item $NNN$ with isospin $I^{(3)} = 1/2$ and spin $S^{(3)} = 1/2$ ($^{3}$H or $^{3}$He).
\end{itemize}

The three-body $\bar{K}NN$ system with different quantum numbers was studied in our previous works,
in particular, quasi-bound state pole positions in the $K^- pp$ system ($\bar{K}NN$ with isospin $I^{(3)} = 1/2$
and spin $S^{(3)} = 0$) were calculated in~\cite{PRC90-II}. The pole positions calculated with coupled $\bar{K}NN$
and $\pi \Sigma N$ channels and $\bar{K}N - \pi \Sigma$ potentials are:
\begin{eqnarray}
\label{zKpp}
\label{SIDD1}
z_{K^-pp}^{\rm SIDD1} &=& -53.3 - i \, 32.4 {\, \rm MeV}, \\
\label{SIDD2}
z_{K^-pp}^{\rm SIDD2} &=& -47.4 - i \, 24.9 {\, \rm MeV}, \\
\label{Chiral}
z_{K^-pp}^{\rm Chiral} &=& -32.2 - i \, 24.3 {\, \rm MeV}, 
\end{eqnarray}
while one-channel calculation of the $\bar{K}NN$ system using exact optical $\bar{K}N(-\pi \Sigma)$ potentials
give slightly different results:
\begin{eqnarray}
\label{zKppOpt}
\label{SIDD1Opt}
z_{K^-pp,Opt}^{\rm SIDD1} &=& -54.2 - i \, 30.5 {\, \rm MeV}, \\
\label{SIDD2Opt}
z_{K^-pp,Opt}^{\rm SIDD2} &=& -47.4 - i \, 23.0 {\, \rm MeV}, \\
\label{ChiralOpt}
z_{K^-pp,Opt}^{\rm Chiral} &=& -32.9 - i \, 24.4 {\, \rm MeV}.
\end{eqnarray}
Comparison of the exact results Eqs.(\ref{SIDD1}--\ref{Chiral}) and the approximated ones
Eqs.(\ref{SIDD1Opt}--\ref{ChiralOpt}) demon\-stra\-tes that using of the exact optical $\bar{K}N(-\pi \Sigma)$
potentials instead of the coupled-channel $\bar{K}N - \pi \Sigma$ ones is quite accurate approximation.

No quasi-bound states caused by pure strong interactions were found in the $K^- d$ system
($\bar{K}NN$ with isospin $I^{(3)} = 1/2$ and spin $S^{(3)} = 1$). 
%The codes for numerical solution
%of the three-body AGS equations for the $\bar{K}NN$ systems were modified to
%perform separabelization of the three-body amplitudes, as described before.
The codes for numerical solution of the three-body AGS equations for the $\bar{K}NN$ systems were modified
to construct separable versions of the three-body amplitudes, as described before.

The three-body AGS equations Eq.(\ref{3AGSsep}) were written and numerically solved for the three-nucleon system
$NNN$ with our new two-term $NN$ potential as an input. The calculated binding energy was
found to be $9.95$ MeV for both ${}^3$H and ${}^3$He nuclei since Coulomb interaction was not taken
into account. 
%The numerical code was afterwards changed for separabelization of the $NNN$ subsystem.
The numerical code was afterwards changed for construction of  separable version of the $NNN$ subsystem.

Finally, the partition of the $2+2$ type $\bar{K}N + NN$ is a system with two non-interacting pairs of particles.
%It was described by special three-body system of AGS equations and separabelized in similar way as $\bar{K}NN$
%and $NNN$ three-body subsystems.
It was described by special three-body system of AGS equations, and its separable version was constructed
in similar way as in $\bar{K}NN$ and $NNN$ three-body subsystems.

\section{Preliminary results and conclusion}

%As the first step we solved the four-body AGS equations for the $K^- ppn$ system, using only one term in
%the separabelization Eq.(\ref{X_EDPE}), which is EDPA method. 
As the first step we solved the four-body AGS equations for the $K^- ppn$ system, using only one separable
term in Eq.(\ref{X_EDPE}), which is EDPA method. 
We used the exact optical versions of two our
phenomenological $\bar{K}N$ potentials with one- and two-pole structure of the $\Lambda(1405)$ resonance
(our chirally motivated model of the antikaon-nucleon interaction has isospin-depen\-dent form-factors,
so that some changes in the four-body program code are necessary). Very preliminary results
for the pole position of the quasi-bound state in the $K^- ppn$ are:
\begin{eqnarray}
\label{zKppn}
z_{K^-ppn,Prelim}^{\rm SIDD1} &=& -75 - i \, 14 {\, \rm MeV} \\
z_{K^-ppn,Prelim}^{\rm SIDD2} &=& -71 - i \, 8 {\, \rm MeV}.
\end{eqnarray}
As in the case of the $K^- pp$ system, the quasi-bound state calculated using one-pole phenome\-nological
$\bar{K}N$ potential (SIDD1) is deeper then that one calculated with the two-pole $V_{\bar{K}N}$, and
it has larger width. The results differ from those shown at the conference since some bugs were found
in the program code after the talk.

To conclude, the four-body Faddeev-type AGS equations for search of the quasi-bound state in
the $\bar{K}NNN$ system were written down.  The code for numerical four-body calculations was written,
preliminary results for the pole positions were obtained. Tests and checks of the program code are still
in progress.

\section*{Acknowledgments}
The author is thankful to the Academy of Sciences of the Czech Republic for provided subsidy
for DSc. researchers. 
The work was supported by GACR grant 19-19640S.

%%%%%%%%%%%%%%%%%%%%%%%%%%%%%%%%%%%%%%%%%%%%%%%%%%%%%%%%%%%%%%

% TODO:
% Provide your bibliography here. You have two options:

% FIRST OPTION - write your entries here directly, following the example below, including Author(s), Title, Journal Ref. with year in parentheses at the end, followed by the DOI number.
%\begin{thebibliography}{99}
%\bibitem{1931_Bethe_ZP_71} H. A. Bethe, {\it Zur Theorie der Metalle. i. Eigenwerte und Eigenfunktionen der linearen Atomkette}, Zeit. f{\"u}r Phys. {\bf 71}, 205 (1931), \doi{10.1007\%2FBF01341708}.
%\bibitem{arXiv:1108.2700} P. Ginsparg, {\it It was twenty years ago today... }, \url{http://arxiv.org/abs/1108.2700}.
%\end{thebibliography}

%%%%%%%%%%%%%%%%%%%%%%%%%%%%%%%%%%%%%%%%%%%%%%%%%%%%%%%%%%%%%%%
%\begin{thebibliography}{00}  %for 2 digits

%%%%%%%%%%%%%%%%%%%%%%%%%%%%%%%%%%%%%%%%%%%%%%%%%%%%%%%%%%%%%%%

% \bibliographystyle{SciPost_bibstyle} % Include this style file here only if you are not using our template
%\bibliography{SciPost_Example_BiBTeX_File.bib}

\nolinenumbers

\end{document}